\pdfoutput=1
\documentclass[11pt]{article}

\usepackage[]{EMNLP2022}

\usepackage{graphicx}
\usepackage{times}
\usepackage{latexsym}

\usepackage[T1]{fontenc}
\usepackage[utf8]{inputenc}

\usepackage{microtype}

\usepackage{inconsolata}

\title{Decomposing the Fundamentals of Creepy Stories}

\author {
    Sakshi Goel, Haripriya Dharmala, \and Yuchen Zhang\\
    University of Southern California \\
    \{sakshigo, dharmala,yzhang71\}@usc.edu
\AND
Keith Burghardt\\
USC Information Sciences Institute\\
keithab@isi.edu
}
\begin{document}

\maketitle
\begin{abstract}
Fear is a universal concept; people crave it in urban legends, scary movies, and modern stories. Open questions remain, however, about why these stories are scary and more generally what scares people. In this study, we explore these questions by analyzing tens of thousands of scary stories on forums (known as subreddits) in a social media website, Reddit. We first explore how writing styles have evolved to keep these stories fresh before we analyze the stable core techniques writers use to make stories scary. We find that writers have changed the themes of their stories over years from haunted houses to school-related themes, body horror, and diseases. Yet some features remain stable; words associated with pseudo-human nouns, such as clown or devil  are more common in scary stories than baselines. In addition, we collect a range of datasets that annotate sentences containing fear. We use these data to develop a high-accuracy fear detection neural network model, which is used to quantify where people express fear in scary stories. We find that sentences describing fear, and words most often seen in scary stories, spike at particular points in a story, possibly as a way to keep the readers on the edge of their seats until the story's conclusion. These results provide a new understanding of how authors cater to their readers, and how fear may manifest in stories. 
\end{abstract}

\section{Introduction}
%{\color{red} [Add robustness checks with \% words in scary stories vs confession alone, TIFU alone, etc.,]}
``The story had held us, round the fire, sufficiently breathless, but except the obvious remark that it was gruesome...I remember no comment uttered till somebody happened to say that it was the only case he had met in which such a visitation had fallen on a child.'' This is the opening line to Henry James's ``The Turning Of The Screw'' \citep{James2008} and summarizes a key and counter-intuitive aspect of scary stories: they can be both revolting and captivating; they capture a fundamental emotion, fear \citep{Adolphs2013}, but in doing so, provide entertainment. The closest analogy might be how spicy foods cause pain yet are tasty. We understand what makes food spicy, and can tailor a meal's spice to fit one's palate, yet it is less clear how to generate fear in a story because we do not know the ingredients that make a story scary. This is a special case of a more general issue, in which we do not yet fully understand how stories elicit particular emotions, i.e., what emotion the reader should feel, in contrast to detecting the emotion a writer expresses \citep{Alhuzali2021,Strapparava2008}.%), which is an under-explored problem in natural language processing.

In this paper, begin to understand how stories make readers scared by quantifying patterns in scary stories. First, we analyze the topics associated with scary stories, and find writers are making fewer stories about houses and more stories about human bodies, school, and diseases, potentially to stay topical. We then develop a method to quantify more universal patterns in scary stories by measuring the percentage of times a word appears in scary and non-scary stories, a metric which we call Scary Story Token Prevalence (SSToP). Words with high SSToPs, such as \emph{creature} or \emph{silhouette}, are more likely to be human-like words, suggesting scary story writers intentionally use words that are semi-human. Finally, we analyze where writers use high SSToP words or words expressing fear within a story. We find that stories can be decomposed into modes where the high SSToP or fearful words appear in regular intervals, and combinations of these modes are used to develop scary story patterns. Overall, these results lend insight into how stories express fear, and patterns writers exploit to make scary stories.

Our contributions can be summarized as follows:
\begin{itemize}
    \item we analyze dynamic patterns of scary stories to understand what themes have become popular or fell out of favor.
    \item we analyze static patterns, such as the relationship between scary story nouns like ``human'' to capture what words writers use and why they might use those words
    \item we capture how scary stories are written by quantifying the common scary story modes. We find some modes are particularly common, which offers clues about how writers make a scary story.
\end{itemize}

\section{Related Work}

\subsection{Uncanny Valley}
What makes something scary is most similar to previous work on the uncanny valley effect. Research on the uncanny valley began with Jentsch and Freud who coined the term \emph{unheimlich} (literally, ``unhomely'') when referencing human uncanniness and eeriness \citep{Freud1909,Jentsch1906}. The uncanny valley also came into prominence among human-computer interaction (HCI) and human-robot interaction (HRI) researchers with a paper by Mori (\citeyear{MoriUncannValley}). Researchers were inspired by Mori to learn what can make a robot unsettling \citep{Broadbent2017} because willingness to work with robots can depend on their likability \citep{Destephe2015}. An objective definition of uncanny (what makes something human-like) or valley (what is and is not unsettling) is still, however, a fiercely debated topic \citep{Mathur2016,Langer2018,Mathur2020,Palomaki2018,MacDorman2006,Ho2017}. Moreover, the qualitative shape of the uncanny valley varies significantly between studies \citep{Mathur2016,Langer2018,Mathur2020,Palomaki2018,MacDorman2006,Ho2017}. Finally, these works (with rare exception \citep{McAndrew2016}) do not explore non-image media, such as text. Open questions, such as how to make less-scary text, or how make scarier ghost stories, have also been under-explored.

\subsection{Emotion Detection}
The fear detection model we create is related to previous work on emotion detection. Emotions are fundamental to humans across cultures \citep{Plutchik1982}, and are typically split into eight components: terror (fear), admiration, ecstacy, vigilance, rage, loathing, grief, and amazement. Research on detecting these emotions in text has a long history \citep{Acheampong2020,Chen2022}, but this research recently came to prominence with text embedding-based methods that can accurately detect emotion in sentences. Detection methods have recently exploited advances in Bi-LSTMs and attention for higher accuracy (c.f., \citep{Cai2018,Alhuzali2021}) and have proven useful for chatbots, for example, which are used in a range of applications \citep{Chen2022}. Emotion detection is closely related to sentiment analysis, such as VADER \citep{Gilbert2014}, or the hedonometer, studied by Dodds et al., \citep{Dodds2011,Dodds2016}. Both methods aim to detect the positivity or negativity of words or sentences, and has been used to understand story arcs \citep{Dodds2016} or tweet sentiment \cite{Dodds2011,Gilbert2014}. %Our manuscript compliments this work by quantifying story arcs for scary words instead of positive or negative words. 

%In the present paper, we focus on how stories are meant to elicit fear, in contrast to emotion detection, which quantifies what emotions are expressed in a sentence (we explore what scares readers rather than whether the text says ``I am scared''). Detecting the emotion the reader should feel is a ripe field of study. It is unknown, for example, the components of a text that can elicit rage or loathing in readers. Instead, to paraphrase a Supreme Court opinion \citep{gewirtz1995know}, we only know it when we see it.

\section{Data}

\subsection{Fear Training Data}
We use multiple datasets to train a fear-detection model, where each dataset has sentences labeled as expressing fear or not. The datasets were gathered and converted into a binary classification problem with a label 1 or 0 representing if the sentence does or does not express fear, respectively. The datasets used for modeling are described in Table~\ref{tab:tab1}. This is one of the most complete collections of data to analyze fear in text. Alike to previous research \citep{Bostan2018}, we collect a range of emotion data, but in contrast to that work, some of our data are unique and only available upon request from the authors \citep{Szpakowicz2007}. This diverse dataset is trained with a model we describe in the Metrics section. In total, there are 70K labeled sentences, with roughly 7K of these sentences labeled ``fear.'' 

\begin{table*}[]
    \centering
    \begin{tabular}{|p{0.4\textwidth}|p{0.5\textwidth}|}\hline
    Dataset	& Description  \\\hline
ISEARs \citep{ISEARS}	& Obtained from cross‐cultural studies in 37 countries and contains sentences annotated for joy, sadness, fear, anger, guilt, disgust and shame emotions \\\hline%& 	Sentences marked ‘fear’ labelled 1, all others labelled 0.\\\hline
WASSA 2017 \citep{WASSA2017}& 	Constructed from tweets and annotated for joy, sadness, fear, and anger emotions\\\hline
Affect Data \citep{Alm2008}& 	Constructed from Tales and classified into angry, fearful, happy, sad, disgusted and surprised emotions\\\hline%	& Sentences marked 3 (data’s label for fear) labelled 1.Remaining labels labelled 0. \\\hline
Emotion Stimulus No Cause Dataset \citep{Ghazi2015}& 	Data developed from FrameNets' annotated data for emotion lexical unit, each sentence is marked with an emotion tag.	\\\hline%& Sentences marked ‘fear’ labelled 1.Sentences marked ‘anger’, ‘sad’, ‘happy’ labelled 0.\\\hline
CrowdFlower Twitter Sentiment Analysis Emotion Data \citep{Crowdflower}& 	Constructed from tweets and classified into 13 emotions (sadness, relief, anger, worry, enthusiasm etc.)\\
\hline%\hline%& 	Sentences marked ‘happiness’, ‘love’, ‘fun’ and ‘enthusiasm’ labelled 0.\\\hline
Emotion-Annotated Dataset \citep{Szpakowicz2007}& 	Constructed from blog data sentences classified into happiness, sadness, fear, anger, surprise and disgust. 	\\\hline%& Sentences marked ‘fear’ labelled 1.Remaining sentences are labelled 0.\\\hline
DailyDialog Dataset \citep{Li2017}	& Human-written text sentences tagged with emotions and communication intention\\\hline%	& Sentences marked 3 (data’s label for fear) labelled 1.Remaining sentences are labelled 0.\\\hline
Emotion Hashtag Corpus \citep{Mohammad2015}&	Tweets annotated with emotions based on the emotion-word hashtag in the tweet text \\\hline%& 	Sentences marked ‘fear’ labelled 1.Remaining sentences are labelled 0.\\\hline
    \end{tabular}
    \caption{Datasets used to construct fear model.}
    \label{tab:tab1}
\end{table*}

\begin{figure*}[tbh!]
    \centering
    \includegraphics[width=\linewidth]{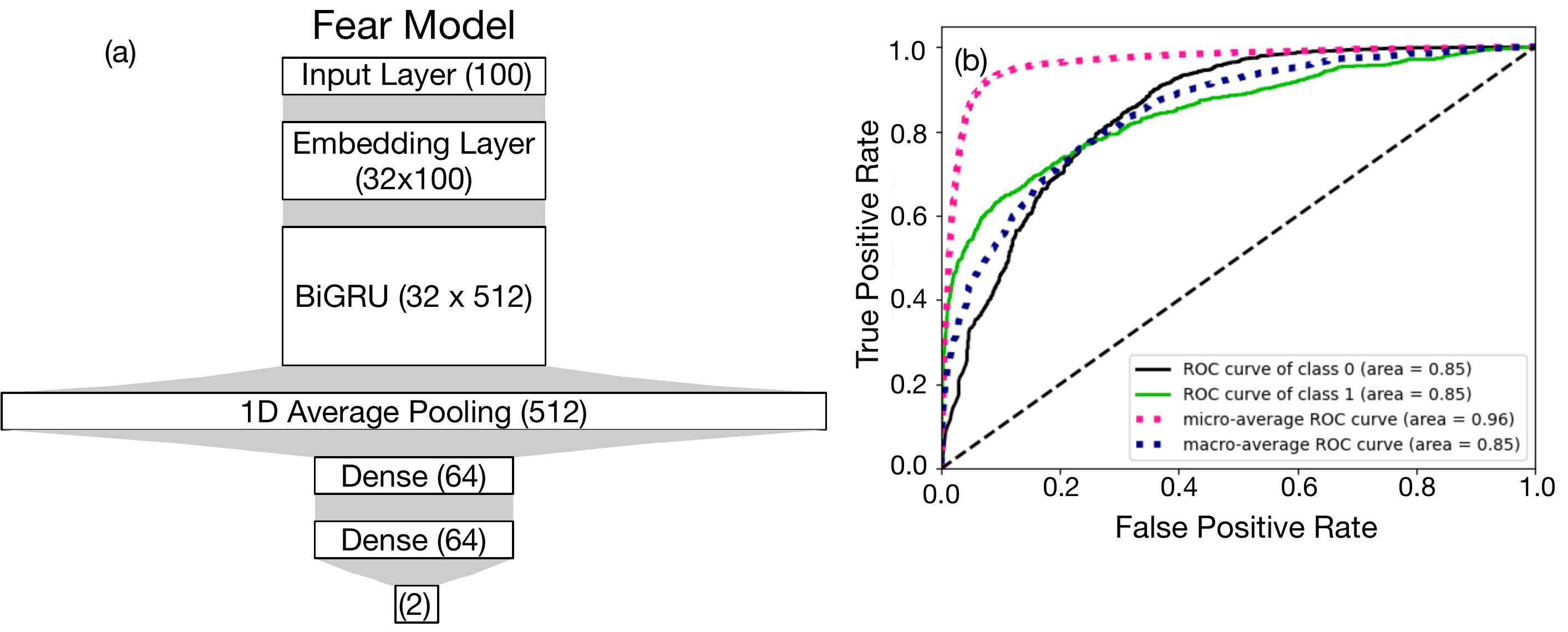}
    \caption{The fear model framework. (a) The neural network used to predict fear in sentences (5,840,850 parameters total). (b) The receiver operative characteristic for the neural network model.}
    \label{fig:fig1}
\end{figure*}
\subsection{Reddit Data}
We extract stories from subreddits using the Python wrapper for the Pushshift API (\url{https://psaw.readthedocs.io}). Stories are collected within two broad genres: scary stories and non-scary stories. For the scary stories, our data comes from the subreddits r/CreepyPasta, r/ShortScaryStories, and r/NoSleep, and for non-scary stories we selected the subreddits r/TIFU, r/Self, r/StoriesAboutKevin, r/Confession, and r/Confessions, which are common story-writing subreddits. All stories are collected from the start of the subreddit until June, 2020. These data are used to train FastText embeddings \citep{joulin2016fasttext}, and to find SSToP scores. After cleaning, described below, there are in total 123K stories.

\subsection{Data cleaning}
To clean data for fear detection, we remove all characters except English alphabets and spaces. This ensures that the fear model’s vocabulary is only English words, and any numbers/HTML markup is removed. Next, we split phrases written in the camel case (a frequent occurrence in the Reddit and Tweet datasets due to hashtags) into its constituent words. (e.g. \#EndThisTyranny split to 'end this tyranny'). Finally, we convert all strings to lowercase and remove all English stopwords (collected from NLTK corpus), to ensure that the model doesn’t falsely learn fear from universally occurring stopwords.  %To calculate the relative word frequency metric, SSToP, we remove stories that have fewer than 500 words in order to create more accurate story modes, described later.% and afterwards we remove words that appear less than 100 times. This reduces noise when calculating the metric. 

\subsection{Metrics}

%\begin{figure*}[tbh!]
%    \centering
%    \includegraphics[width=\linewidth]{figures/Fig1.pdf}
%    \caption{The uncanny valley. (a) A schematic of the uncanny ``valley,'' where higher values are more scary, in order to better align with Fig.~\ref{fig:fig4}. (b) The fear detection neural network architecture. (c) The scariness metric.}
%    \label{fig:fig1}
%\end{figure*}
\subsubsection{Fear Confidence}
%\begin{figure*}[tbh!]
%    \centering
%    \includegraphics[width=0.7\linewidth]{figures/FigMethods1.pdf}
%    \caption{Fear model performance. (a) Accuracy for training and validation sets versus model epoch. (b) ROC curves for the test set.}
%    \label{fig:fig2}
%\end{figure*}
To determine the confidence that a word or sentence expresses fear we train a model whose architecture is described in Fig.~\ref{fig:fig1}a. A Bidirectional GRU (Bi-GRU) is used because it is capable of incorporating contextual information in the forward and reverse direction in contrast to a standard GRU or LSTM. A Global Average Pooling layer was used after the Bi-GRU layer and a dropout layer of 50\% was used to regularize the dense layer just before the two-node prediction layer. A batch size of 64 was used with categorical cross-entropy loss and an Adam optimizer. %The output of the model is the confidence that the sentence has fear, an example of these confidences are shown in Fig.~\ref{fig:fig1}b. 

To train this model, we split the meta-dataset into 3 parts (where each datapoint is assigned at random): 75\% for training, 10\% for validation and 10\% for testing. The original training dataset contained 55.8K sentences, of which 5.5K were sentences expressing fear. To reduce imbalance, we up-sample the fear sentences to create 100.5K sentences for training. Validation and test sets are separate sets of 7.0K sentences without up-sampling or down-sampling (roughly 700 sentences in each set were labeled \emph{fear}). Tensorflow Keras is used for the tasks of tokenization, sequencing and padding of the three datasets. The tokenizer is only fit on the training data. The vocabulary size was 56.5K. Each sentence is then converted into a sequence and zero-padded to a length of 100. The model ROC curve is shown in Fig.~\ref{fig:fig1}b, which shows high micro- and macro-averaged AUC (0.96 and 0.85, respectively).

We compare this model to logistic regression models  with L1 or L2 regularization. Text is tokenized and fed through FastText \citep{joulin2016fasttext} embeddings trained on the scary and non-scary story corpora. All embedded words in a story are then averaged to create a document embedding. When training the logistic regression models, the training data is down-sampled so that each class is of equal size. The comparisons against the baselines are shown in Table~\ref{tab:baselines}. We see significant improvements over the baseline models. %The ROC curve, as well as training and validation accuracy versus epoch can be seen in Fig.~\ref{fig:fig1}. The distribution of values for words is seen in Fig.~\ref{fig:fig1}b. The words are of those in the FastText dictionary.

\begin{table*}[]
    \centering
    \begin{tabular}{|l|c|c|c|}
    \hline
         Model & Accuracy & F1 & ROC-AUC\\
         &                & (Macro Avg) &\\\hline
         {\bf Neural Net} &\textbf{0.92} &\textbf{0.76}&\textbf{0.85}\\
         Logistic (L1) & 0.71&0.60&0.84\\
         Logistic (L2) & 0.74&0.61&0.84\\\hline
         Majority & 0.90 & 0.57&0.5\\\hline
    \end{tabular}
    \caption{Model performance for neural network-based fear model and logistic regression baselines. Bottom row are metrics for majority class null model, in which we guess all test data is not scary (the majority class).}
    \label{tab:baselines}
\end{table*}

\subsubsection{Scary Story Token Prevalence}

SSToP is defined as the odds ratio between a word's frequency in scary and non-scary stories
\begin{equation}\label{eq:eq1}
    c(word) = \frac{f_{scary}(word)}{f_{normal}(word)}
\end{equation}
Where $f_X(word)$ is the word frequency across all posts and subreddits within the genre $X$. To check the robustness of this method, we re-calculate $c(word)$ such that non-scary word frequencies were calculated from each of the non-scary subreddits (creating five separate metrics) and results are qualitatively the same. 
\begin{table*}[tbh!]
    \centering
    \begin{tabular}{|l|c||l|c|}
    \hline
\textbf{High SSToP Words} & \textbf{SSToP} &\textbf{Low SSToP Words} & \textbf{SSToP}\\
\textbf{(Common in Scary Stories)} &&\textbf{(Uncommon in Scary Stories)}&\\\hline
    eyeless	& 100&rself&	0.0018\\
scarecrows	& 108&fwb	&0.0021\\
cryptids&	110&superjail&	0.0022\\
skinwalkers&	130&19f	&0.0023\\
headstones&	137&tldr&0.0025\\
    \hline
    \end{tabular}
    \caption{Examples of words with high and low SSToP scores. Skin-walkers, which have a high SSToP score, are a harmful witch in Navajo culture.}
    \label{tab:tab3}
\end{table*}
To calculate this metric for each word, we first select all stories which contain at least 500 words. This helps ensure that the lengths of the stories are approximately uniform and allows us to calculate story modes described later. The stories are preprocessed by tokenizing and removing punctuation. We utilize a dictionary to keep track of the number of times each word occurs. This is calculated for both scary and non-scary stories and Eq.~\ref{eq:eq1} is applied to determine a SSToP value for each word. Since Reddit stories are not checked for correct grammar or spelling, there are several unique names or words which may be misspelt. Because the occurrence of each of these words does not span broadly throughout the entire subreddit, SSToP cannot accurately assign a value to these unique terms. Therefore, we only consider words which have occurred at least 100 times in the entire corpora of scary and non-scary stories. %Finally, the hedonometer is represented as a dictionary of unique words and their SSToP values. 

\subsection{Generating Story Modes}
The main motivation behind creating story modes is to identify patterns writers exploit to make scarier stories. Some stories may, for example, get scary towards the end while other stories may start out scary and end scary, but have a less-scary segment in the middle. Therefore, while we generate a model to detect whether fear is expressed in sentences, it is the pattern of fear, the modes, that we study to better understand how readers might be provoked to experience fear.

To construct these story modes, we first segment each story into ten points based on the percentage of story covered (0\%, 10\%,...,90\%). We then calculate the fear confidence of a sentence located at that point in the story or average the SSToP score for 50 words after that point in the story. After the metrics are sampled, we construct a matrix, $\mathbf{M}$, such that columns represent the 10 sampled points, and rows represent each story.  We then apply singular value decomposition to calculate the story modes \citep{Dodds2016}, such that $\mathbf{M} = \mathbf{V\Sigma U} = \mathbf{WU}$. The matrix $\mathbf{U}$ represents the modes while $\mathbf{W}$ represents the coefficients of each mode used to reconstruct each story, alike to the coefficients of a Fourier transform. The largest element in each row of $\mathbf{W}$ represents the mode that best aligns with each story, but in general several modes are needed to reconstruct a story. 

\subsection{Word similarity to human}
To determine how similar a noun is to \emph{human}, we first embed nouns using FastText. Human synonyms, namely human, mortal, person, soul, Homo Sapien, earthling, higher animal, and living person (common synonyms of human that are gender neutral) are averaged together into one \emph{human} vector to avoid noise associated with the embedding of a single word. This follows similar work on gender bias in word embeddings, where the first principle component of a collection of words are used to construct a he-she vector, in order to reduce noise associated with each individual pair of words \citep{Bolukbasi2016}. Next we determine what words in the word embeddings are nouns using Python's TextBlob library (\url{https://textblob.readthedocs.io}). If NN, PN, or NP are in the TextBlog tags for a word, the word is categorized as a noun. This method helps us avoid finding the unrealistic similarity between, e.g., a verb and a noun. Finally, we determine distance between words using Euclidean distance. Other distance metrics, such as Manhattan distance \citep{Aggarwal2001}, show similar results.

\section{Results}

\begin{figure*}[tbh!]
    \centering
    \includegraphics[width=\linewidth]{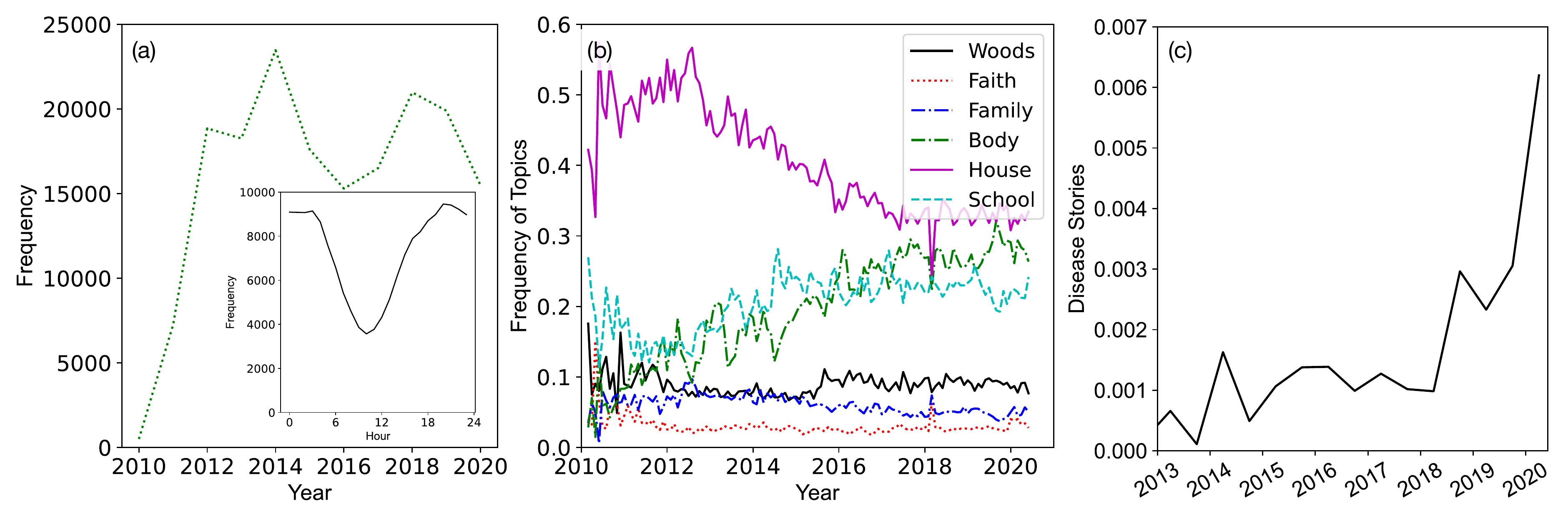}
    \caption{Scary story statistics. (a) The number of stories each year (inset: the number of stories written each hour in GMT time). (b) LDA topics over time, which shows a significant drop in the house-like topic but an increase in body-related and school-related topics in recent years. (c) The proportion of disease-related stories over time.}
    \label{fig:fig2}
\end{figure*}

We show data statistics in Fig.~\ref{fig:fig2}. This figure demonstrates when writers post their stories and what they discuss. First, Fig.~\ref{fig:fig2}a shows the frequency of stories over time, which rapidly increases to over 20,000 stories a year before dropping slightly. In the inset, we see that users are least likely to post their stories at 12 p.m. GMT, or around 5 a.m. EST, perhaps because most writers might be from America and do not post when they just wake up. Next, we use Latent Derichlet Allocation (with lemmatized words and all stop words removed) to find ten story topics, the largest of which are shown in Fig.~\ref{fig:fig2}b. We notice that the house topic (with words like ``door" or ``window'') drops off in frequency while topics related to bodies (with words like ``bodies" or ``blood'') and school (with words like ``phone'' or ``school'') become increasingly popular. These results suggest long-term trends in what writers consider topical for their stories, possibly because writers focus on different topics as they age or they believe readers find school or body horror more interesting. Finally, we show the frequency of disease-like stories, in which ``lockdown,'' ``infect,'' ``viru,'' or ``diseas'' lemmas are within the story, increase dramatically in 2020 (likely due to COVID-19 being in the news), but this follows a long-term trend, in which disease-like stories have become increasingly popular. We also notice a small peak in their prevalence in mid-2014, when the disease Ebola became topical. 

\begin{figure}
    \centering
    \includegraphics[width=0.75\linewidth]{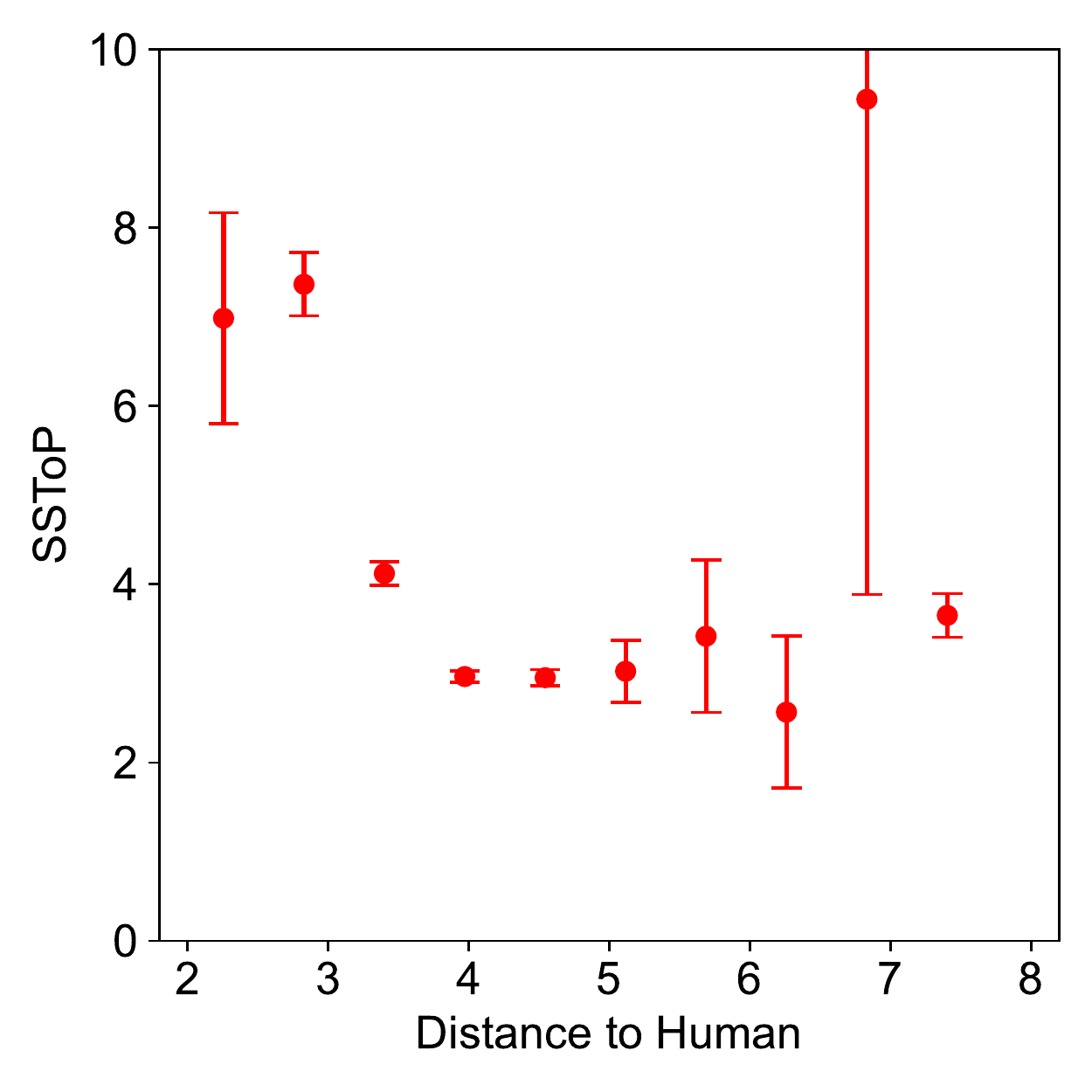}
    \caption{Scary story writers use more human-like words. Mean SSToP score (fraction of words in scary stories) in red versus noun similarity to ``human.'' Error bars are standard errors.}
    \label{fig:fig3}
\end{figure}

Next, we show in Fig.~\ref{fig:fig3} that a word's SSToP score varies with its similarity to human. Words seen in scary stories tend to be more similar to human, which is reminiscent of the uncanny valley, in which people find semi-human objects more unsettling. A number of plausible reasons for these patterns exist. For example, scary story authors use semi-human words (such as ``devil'' or ``clown'') in an attempt to unsettle readers, or they are discussing human-related topics, or they avoid words that are static and boring, such as cars, which are far from ``human'' in latent embedding space. Whatever the reason, this pattern gives new insights into how writers make scary stories.

\begin{figure*}[tbh!]
    \centering
    \includegraphics[width=0.95\linewidth]{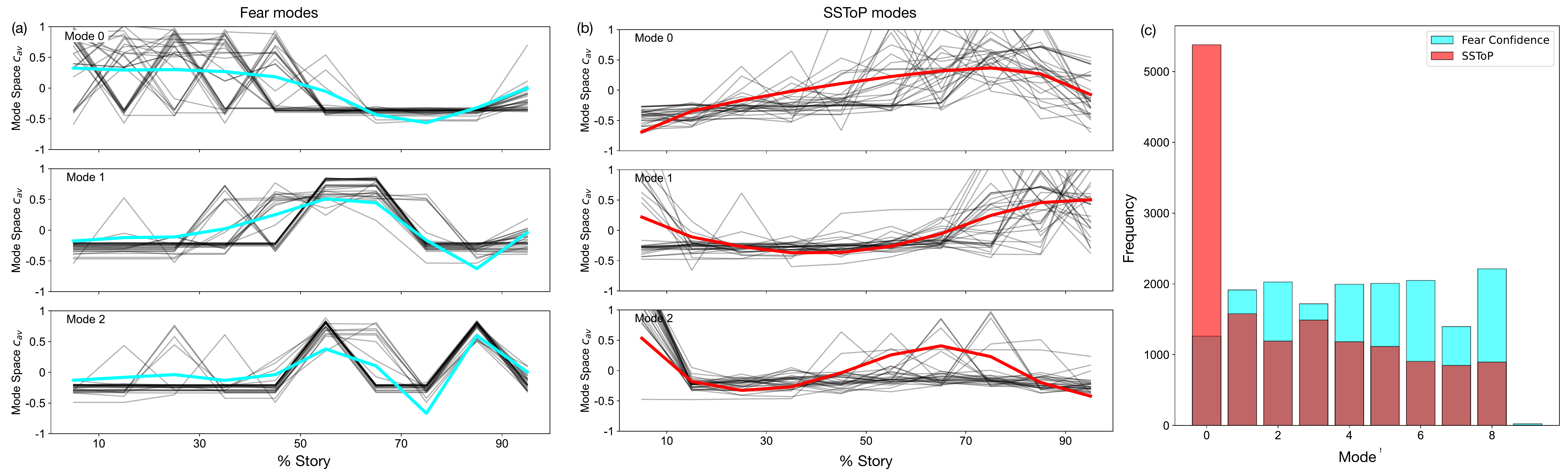}%Fig3.pdf}
    \caption{Story modes in r/NoSleep. (a) 3 lowest story modes for fear (modes are blue lines), (b) 3 lowest story modes for SSToP scores (modes are red lines), and (c) histograms of each story mode for the respective metrics. Gray lines in (a-b) are example stories that best match those respective modes (they have the largest W coefficients for those modes). %This closely follows the methodology seen in Reagan et al. (\citeyear{Dodds2016}).
    }
    \label{fig:fig4}
\end{figure*}
Finally, we analyze how scary story authors develop story modes. In Fig.~\ref{fig:fig4}, we show story modes of fear confidence and SSToP score in the r/NoSleep subreddit (14K stories in total). Blue lines in Fig.~\ref{fig:fig4}a correspond to the three lowest modes for fear confidences, red lines in Fig.~\ref{fig:fig4}b are the three lowest modes for SSToP scores. Gray lines correspond to 20 stories which have the highest coefficients in the corresponding $\mathbf{W}$ matrix for a given mode. We normalize the story modes by dividing by the largest $\mathbf{W}$ coefficient in each row. These figures show that stories use high SSToP words, such as ``shadow'' or ``creature'' at very particular points in the story, presumably at story climaxes or key scenes. Stories also express fear at particular points in the story as well, presumably to keep the audience thrilled or afraid until the end of the story.

In Fig.~\ref{fig:fig4}c, we show the frequency of the largest mode associated with each story. We see that many stories are in the lowest SSToP mode. Scary story authors therefore tend to use scary words in the center of the story, perhaps to set up scenes. Surprisingly, however, there are a variety of modes used for fear, and a combination of these modes often appear in stories. Furthermore, the fear and SSToP modes rarely align. Namely, the largest SSToP modes for a given story have a low correlation with the top fear modes (Spearman correlation $=0.06$, p-value $<10^{-10}$). Finally mode 9 is rarely the largest mode because it corresponds to a story without any change (the mode is a flat line).

\section{Conclusions}

In conclusion, we analyze text patterns to better understand how writers make scary stories. The results suggest that writers tend to be topical or change the story topic to fit the mood of readers. Moreover, scary story authors tend to use human-like words, which is reminiscent of the uncanny valley in which people find human-like images more unsettling. In addition, authors create stories whose modes may help scare readers. Overall, our work uncovers patterns associated with text that elicits fear, but we believe more work will be needed to understand how these patterns may cause fear to be elicited.

\section{Limitations}
There are a number of limitations in our work. While we uncover patterns associated with scary stories, more work is needed to understand whether or how these patterns create fear in readers. This task will require analyzing reader reactions in a controlled study. In addition, the relationship between nouns and words like ``human'' have not been tested on alternative text embedding models, such as GLoVe \citep{pennington2014glove}, or attention-based embeddings, such as BERT \citep{Devlin2019}. Finally story motifs we uncover should be compared against alternative methods \citep{Dodds2016}. 

\section{Ethical Considerations}
All results come from publicly available and anonymized data, therefore the study poses minimal ethical risk.  

\section{Acknowledgements}
This project was developed at the USC Center for Knowledge-Powered Interdisciplinary Data Science (CKIDS) DataFest 2020.  We also thank the USC Graduates Rising in Information and Data Science (GRIDS) for their help.  Code to reproduce some of our major results are shown here: https://github.com/KeithBurghardt/Creepypasta.

%\bibliography{references}
%\bibliographystyle{acl_natbib}

%\appendix

%\section{Example Appendix}
%\label{sec:appendix}

%This is an appendix.

\end{document}